\documentclass[aps,prl,twocolumn,superscriptaddress,showpacs,10pt]{revtex4-1}

\usepackage{amsmath,amssymb,bm}
\usepackage{graphicx}

\begin{document}

\title{Cavity-loss induced plateau in coupled cavity QED array}

\author{Tatsuro Yuge}
\affiliation{Department of Physics, Osaka University, Machikaneyama-Cho, Toyonaka, Osaka, 560-0043, Japan}

\author{Kenji Kamide}
\affiliation{Institute of Industrial Science, the University of Tokyo, Komaba, Meguro-ku, Tokyo, 153-8505, Japan}

\author{Makoto Yamaguchi}
\affiliation{Department of Physics, Osaka University, Machikaneyama-Cho, Toyonaka, Osaka, 560-0043, Japan}

\author{Tetsuo Ogawa}
\affiliation{Department of Physics, Osaka University, Machikaneyama-Cho, Toyonaka, Osaka, 560-0043, Japan}

\date{\today}

\begin{abstract}
Nonequilibrium steady states are investigated in a coupled cavity QED array system 
which is pumped by a thermal bath and dissipated through cavity loss. 
In the coherent (non-zero photon amplitude) phase, plateau regions appear, 
where the steady states become unchanged against the variation of the chemical potential of the thermal bath. 
The cavity loss plays a crucial role for the plateaus: 
the plateaus appear only if the cavity loss exists, 
and the photon leakage current, which is induced by the loss, is essential to the mechanism of the plateaus. 
\end{abstract}


\maketitle

{\it Introduction}.---In recent years, much attention has been paid to coupled cavity QED arrays. 
In these systems, the effective photon-photon interaction in each cavity and photon hopping between cavities 
lead to rich quantum many-body phenomena of photons \cite{Review1,Review2}. 
For example, in equilibrium states, these systems exhibit quantum phase transitions 
between the ``incoherent (or Mott insulator)'' and ``coherent (or superfluid)'' phases 
\cite{Greentree_etal,Hartmann_etal,Angelakis_etal,RossiniFazio,Makin_etal,KochHur,Schiro_etal,Kamide_etal}. 
Nonequilibrium states in these systems have been also investigated 
\cite{Carusotto_etal,Hartmann,Tomadin_etal,Ferretti_etal,Schmidt_etal,Knap_etal,Nissen_etal,Grujic_etal,Jin_etal}. 
Most of the works have focused on the nonequilibrium effects of the pumping and dissipation 
on the properties observed in equilibrium states; e.g., signatures of the two phases in nonequilibrium states.   
However, phenomena genuinely characteristic to nonequilibrium states in the systems remain unexplored. 


Apart from the coupled cavity QED array, 
examples of this kind of phenomena are seen in ``nonequilibrium phase transitions,'' such as 
the pattern formation outside equilibrium \cite{CrossHohenberg}, 
dissipative structure \cite{NicolisPrigogine}, 
and lasing phenomena \cite{Haken}. 
These are observed not in equilibrium states but only in nonequilibrium states. 
It is important to find such phenomena 
from the viewpoint that they can be used as test models for nonequilibrium statistical mechanics and thermodynamics. 

In this paper, we report that another genuine nonequilibrium phenomenon can occur in a coupled cavity QED array 
that is pumped by a thermal bath and dissipated through cavity loss. 
We find that in the coherent phase there exist plateau regions 
where the states of the array system becomes unchanged to the variation of 
the chemical potential of the thermal bath. 
We show that the plateau emerges only if the cavity loss (dissipation) exists, 
which implies that the nonequilibrium effect is crucial for this phenomenon.

\begin{figure}[t]
\begin{center}
\includegraphics[width=\linewidth]{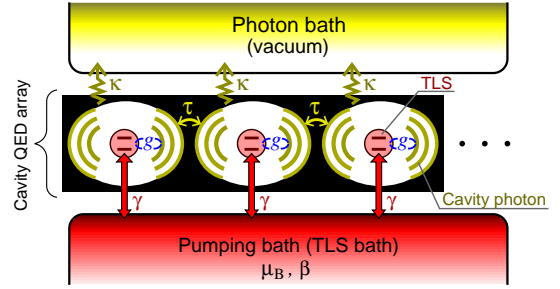}
\end{center}
\caption{(Color online) An illustration of the setup, 
consisting of a cavity QED array, a pumping bath 
(equilibrium state with the chemical potential $\mu_{\rm B}$ and inverse temperature $\beta$), 
and a photon bath (vacuum state). 
Each cavity photon couples with a two-level system (TLS) (coupling constant $g$), 
with neighboring cavity photons (hopping constant $\tau$), 
and with the photon bath (which induces the cavity loss $\kappa$). 
Each TLS is incoherently pumped (rate $\gamma$) by the pumping bath. 
}
\label{fig:schematic}
\end{figure}

{\it Setup}.---We consider a cavity array that is connected to two baths. 
An illustration of the total system (array + baths) is shown in Fig.~\ref{fig:schematic}. 
Each cavity contains a two-level system (TLS) 
which is continuously excited through the coupling (excitation exchange) with the pumping bath (TLS bath). 
The pumping bath is in an equilibrium state 
characterized by the chemical potential $\mu_{\rm B}$ and inverse temperature $\beta$. 
Thus the TLSs in the cavity array are incoherently pumped (rate $\gamma$). 
The photon mode in each cavity interacts with the TLS in the cavity (coupling constant $g$) 
and is coupled with the photons in the neighboring cavities through mode overlap (hopping constant $\tau$). 
It is also coupled with the photon bath (vacuum), which induces the cavity loss $\kappa$.

The coupled cavity QED array is described by the Jaynes-Cummings-Hubbard Hamiltonian ($\hbar=1$) 
\cite{Greentree_etal}:
\begin{align}
\hat{H}_{\rm S} 
&= \sum_i \hat{H}_i^{\rm JC} - \tau \sum_{\langle ij \rangle} \hat{a}_i^\dag \hat{a}_j, 
\label{H_S}
\\
\hat{H}_i^{\rm JC} &= \omega_0 \hat{\sigma}_i^+ \hat{\sigma}_i^- + \omega_0 \hat{a}_i^\dag \hat{a}_i 
+ g(\hat{\sigma}_i^+ \hat{a}_i + \hat{\sigma}_i^- \hat{a}_i^\dag),
\label{H_JC}
\end{align}
where $i$ and $j$ represent the site indices of the cavities, 
$\hat{\sigma}_i^+$ and $\hat{\sigma}_i^-$ are the raising and lowering operators of the TLS, 
and $\hat{a}_i^\dag$ and $\hat{a}_i$ are the creation and annihilation operators of the cavity photon 
in the $i$th cavity.
The first term in Eq.~(\ref{H_S}) is the Jaynes-Cummings Hamiltonian 
for single cavities given by Eq.~(\ref{H_JC}), 
and the second term is the hopping of the photons between the neighboring cavities 
($\langle ij \rangle$ means the nearest neighbor sites). 
We assume that the transition frequency of the TLS and the cavity frequency are resonant 
and the same for all of the cavities (denoted by $\omega_0$).

To obtain insight over the steady-state properties of the pumped-dissipative system, 
we employ a mean-field (MF) approximation and quantum master equation (QME). 
We introduce the steady-state average photon field $\psi(t) \equiv \langle \hat{a}_i \rangle_{\rm ss}$ as the mean field. 
We assume that it oscillates with a single frequency $\mu$ in the rest frame: 
$\psi(t) = \psi e^{-i\mu t}$, where the amplitude $\psi$ is a real number. 
After using the MF approximation \cite{Greentree_etal} and transforming into the rotating frame with $\mu$, 
we take the trace over the baths and use the Born-Markov approximation. 
Then we have the QME in the Shr\"odinger picture (within the rotating frame) \cite{BreuerPetruccione}: 
$d\tilde{\rho}/dt = \mathcal{L}\tilde{\rho}$, 
where $\tilde{\rho}$ is the density matrix of the array system in the rotating frame. 
The QME superoperator $\mathcal{L}$ is given by 
$\mathcal{L} = -i\bigl[ \tilde{K}_{\rm S}, ~\bigr] + \mathcal{L}_{\rm TLS} + \mathcal{L}_{\rm ph}$, 
where $\tilde{K}_{\rm S}$ is the effective system Hamiltonian in the rotating frame, and 
$\mathcal{L}_{\rm TLS}$ and $\mathcal{L}_{\rm ph}$ are the superoperators 
associated with the TLS and photon baths, respectively. 
We will describe the details in the last part of this paper. 
We determine the steady-state solution $\tilde{\rho}_{\rm ss}(\psi,\mu)$ in the rotating frame by
\begin{align}
\mathcal{L}\tilde{\rho}_{\rm ss}(\psi,\mu) =0. 
\label{steadyStateEq}
\end{align}
Note that the solution $\tilde{\rho}_{\rm ss}$ depends on $\psi$ and $\mu$. 
Then we have the self-consistent equation for the mean field: 
\begin{align}
\psi = {\rm Tr}_{\rm S} [\tilde{\rho}_{\rm ss}(\psi,\mu) \hat{a}].
\label{selfConsistentEq}
\end{align} 
By solving Eqs.~(\ref{steadyStateEq}) and (\ref{selfConsistentEq}) numerically, 
we determine the values of $\psi$ and $\mu$, and obtain the steady-state density matrix $\tilde{\rho}_{\rm ss}$ 
of the MF-QME for the pumped-dissipative cavity QED array system. 
For the numerical calculation we set the cutoff for the maximum photon number in the cavity to 10. 

\begin{figure}[t]
\begin{center}
\includegraphics[width=\linewidth]{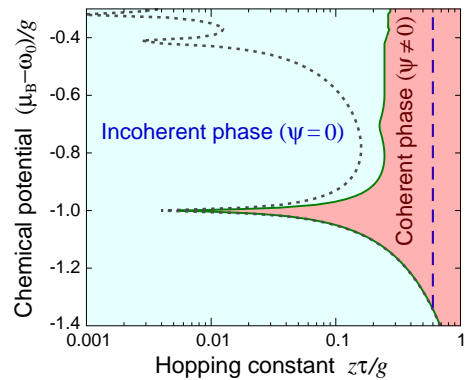}
\end{center}
\caption{(Color online) Steady-state phase diagram of the pumped-dissipative cavity QED array system. 
The vertical and horizontal axes are respectively the chemical potential $\mu_{\rm B}$ of the TLS bath 
and the hopping constant $\tau$ ($z$ is the number of the nearest neighbor sites). 
The boundary between the incoherent ($\psi=0$) and coherent ($\psi\neq 0$) phases is plotted. 
The parameter values are: $1/\beta g = 0.001$, $\gamma/g=0.02$, and $\kappa/g=0.007$.
The dotted curve is the phase boundary for the equilibrium case ($\kappa=0$). 
The dashed line ($z\tau/g=0.6$) is the one along which the results in Fig.~\ref{fig:frequency} are plotted. 
}
\label{fig:phasediagram}
\end{figure}

{\it Phase diagram}.---First we investigate a phase diagram of the nonequilibrium steady states in the array system 
by employing $\psi$ as the order parameter. 
Figure~\ref{fig:phasediagram} depicts the regions for $\psi=0$ and $\psi\neq 0$ 
as a function of the hopping constant $\tau$ and the chemical potential $\mu_{\rm B}$ of the TLS bath. 
We refer to the region with $\psi=0$ as incoherent phase and to the region with $\psi\neq 0$ as coherent phase. 
For comparison, we also show the phase boundary in the equilibrium states ($\kappa = 0$) by the dotted curve. 

In small $\mu_{\rm B}$ (weak pumping) regime, the phase boundary 
is almost the same as that of the equilibrium states. 
This implies that the array system is nearly equilibrium in this regime. 

In larger $\mu_{\rm B}$ (strong pumping) regime, by contrast, 
the coherent phase region shrinks compared with the equilibrium case. 
This results from a nonequilibrium effect; 
in this regime, the number of the cavity photons increases, 
so that the effect of the cavity loss becomes larger. 
Since the loss destroys the coherence of the system, 
the area of the coherent phase decreases in this regime. 

\begin{figure}[t]
\begin{center}
\includegraphics[width=\linewidth]{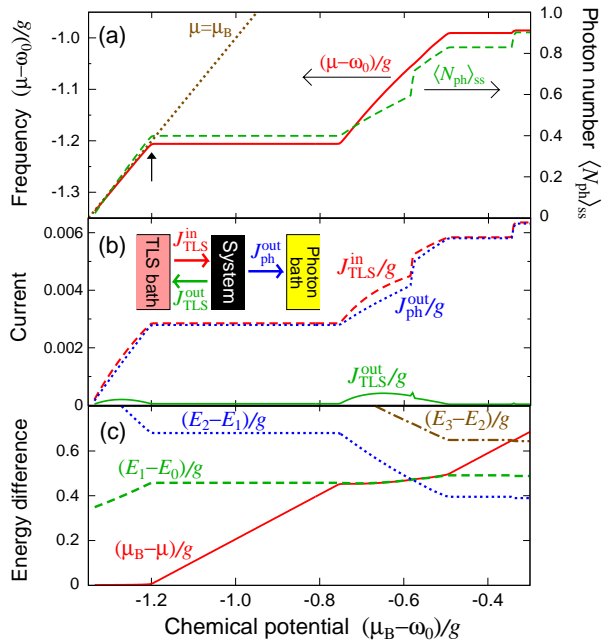}
\end{center}
\caption{(Color online) (a) Photon frequency $\mu$ (solid line; left vertical axis) 
and photon number $\langle N_{\rm ph} \rangle_{\rm ss}$ (dashed line; right vertical axis) 
as functions of the TLS bath chemical potential $\mu_{\rm B}$, 
which are calculated along the dashed line ($z\tau/g=0.6$) in Fig.~\ref{fig:phasediagram}. 
The dotted line represents $\mu=\mu_{\rm B}$. 
(b) Currents between the system and baths. 
The solid, dashed, and dotted lines represent $J_{\rm TLS}^{\rm out}$ (from the system to TLS bath), 
$J_{\rm TLS}^{\rm in}$ (from the TLS bath to system), 
and $J_{\rm ph}^{\rm out}$ (from the system to photon bath), respectively. 
The inset shows an illustration of the currents. 
(c) $\mu_{\rm B}-\mu$ (solid line), $E_1-E_0$ (dashed line), $E_2-E_1$ (dotted line), and  $E_3-E_2$ (dash-dotted line) 
plotted against $\mu_{\rm B}$, 
where $E_q$ ($q=0,1,2,3$) are the eigenenergies of $\tilde{K}_{\rm S}$.
The parameter values are the same as those in Fig.~\ref{fig:phasediagram}. 
}
\label{fig:frequency}
\end{figure}

{\it Main result}.---In Fig.~\ref{fig:frequency}(a), we plot $\mu_{\rm B}$-dependence 
of the frequency $\mu$ and the average photon number per cavity 
$\langle N_{\rm ph} \rangle_{\rm ss} \equiv {\rm Tr}_{\rm S} [\tilde{\rho}_{\rm ss} \hat{a}^\dag \hat{a}]$ 
in the coherent phase along the dashed line in Fig.~\ref{fig:phasediagram} \cite{note1}.
In small $\mu_{\rm B}$ (weak pumping) regime, we observe $\mu\simeq\mu_{\rm B}$. 
We note that in the equilibrium states $\mu=\mu_{\rm B}$ holds. 
Therefore this result is consistent with the fact that the array system is near to equilibrium in this regime. 

In larger $\mu_{\rm B}$ regime, we observe plateaus both in $\mu$ and $\langle N_{\rm ph} \rangle_{\rm ss}$; 
i.e., there are regions of $\mu_{\rm B}$ where $\mu$ and $\langle N_{\rm ph} \rangle_{\rm ss}$ take constant values. 
We numerically confirm that 
the steady-state density matrix $\tilde{\rho}_{\rm ss}$ itself is unchanged in the plateau regions.
We note that the cavity loss is crucial for the emergence of the plateaus 
because $\mu=\mu_{\rm B}$ holds for $\kappa=0$ (in the equilibrium states). 
Furthermore, in the supplement \cite{supplement}, we will show 
that the chemical potential at the beginning of the first plateau [pointed with the up arrow in Fig.~\ref{fig:frequency}(a)] 
increases as $\kappa$ decreases 
and that it enters the unstable region \cite{KochHur} in the limit of $\kappa\to 0$. 
These cavity-loss induced plateaus, which are our main findings in this paper, seem counterintuitive 
since the loss usually smears out (rather than generates) sharp-cut phenomena. 
We also note that the incoherent pumping is necessary to observe the plateaus 
because we see them in the dependence on $\mu_{\rm B}$. 
%
The emergence of the plateaus implies that 
the properties of the cavity QED array system are completely stable to the uncertainty of $\mu_{\rm B}$; 
for example, photoluminescence spectrum and photon statistics 
as well as $\mu$ and $\langle N_{\rm ph} \rangle_{\rm ss}$ 
are unchanged with the variation of $\mu_{\rm B}$ within the plateaus.

{\it Mechanism}.---We here discuss the mechanism of the plateaus. 
For this purpose, we investigate the relationship of  the plateaus with the currents and energy differences. 

As we mentioned, the plateaus are induced by the cavity loss. 
This leads us to expect that quantities directly related to the cavity loss are essential to the mechanism. 
A typical one is the photon current that leaks to the photon bath. 
Therefore we first investigate the relationship with the currents. 

In Fig.~\ref{fig:frequency}(b), we plot the currents: 
$J_{\rm TLS}^{\rm out} 
\equiv -{\rm Tr}_{\rm S} [\hat{\sigma}^+ \hat{\sigma}^- \mathcal{L}_{\rm TLS}^- \tilde{\rho}_{\rm ss}]$ 
(from the system to TLS bath), 
$J_{\rm TLS}^{\rm in} 
\equiv {\rm Tr}_{\rm S} [\hat{\sigma}^+ \hat{\sigma}^- \mathcal{L}_{\rm TLS}^+ \tilde{\rho}_{\rm ss}]$ 
(from the TLS bath to system), and 
$J_{\rm ph}^{\rm out} 
\equiv -{\rm Tr}_{\rm S} [\hat{a}^\dag \hat{a} \mathcal{L}_{\rm ph} \tilde{\rho}_{\rm ss}]$ 
(from the system to photon bath) \cite{note2}, 
where the superoperators $\mathcal{L}_{\rm TLS}^\pm$ and $\mathcal{L}_{\rm ph}$ 
are respectively given by Eqs.~(\ref{L_pump}) and (\ref{L_diss}) (see the last part). 
A schematic illustration of the currents are shown in the inset of Fig.~\ref{fig:frequency}(b). 
We observe that $J_{\rm TLS}^{\rm in} \simeq J_{\rm ph}^{\rm out}$ and $J_{\rm TLS}^{\rm out} \simeq 0$ in the plateaus 
(note that $J_{\rm TLS}^{\rm in} = J_{\rm TLS}^{\rm out} + J_{\rm ph}^{\rm out}$ holds in the steady states).

Next we investigate the relationship with the energy differences. 
As we mentioned, the plateaus are observed in the dependence on the chemical potential $\mu_{\rm B}$. 
In the QME, especially in the superoperator $\mathcal{L}_{\rm TLS}$, 
$\mu_{\rm B}-\mu$ (rather than $\mu_{\rm B}$ itself) 
and the eigenenergy differences appear [see the below of Eq.~(\ref{L_pump})]. 
Therefore we investigate $\mu_{\rm B}-\mu$ and the eigenenergy differences. 

In Fig.~\ref{fig:frequency}(c), we plot $\mu_{\rm B}-\mu$ and $E_q-E_{q-1}$, 
where $E_q$ ($q=0,1,2,3$) are the eigenenergies for the ground and $q$th excited states 
of the effective Hamiltonian $\tilde{K}_{\rm S}$. 
We observe that 
$E_1-E_0$ and $E_3-E_2$ cross with $\mu_{\rm B}-\mu$ respectively at the ends of the first and second plateaus, 
whereas $E_2-E_1$ crosses with $\mu_{\rm B}-\mu$ in the transition region between the first and second plateaus. 



From these observations, we have a following  mechanism scenario of the plateaus. 
First we discuss the first plateau. 
When $\mu_{\rm B}-\mu<E_1-E_0$, i.e., before the end of the first plateau, 
the ground state is the only channel available for pumping from the TLS bath. 
Before the first plateau starts, $J_{\rm TLS}^{\rm in}$ is a convex function of $\mu_{\rm B}$, 
whereas $J_{\rm ph}^{\rm out}$ is almost proportional to $\mu_{\rm B}$. 
These behaviors of the currents possibly result from the fact that, 
for each single pumping channel, $\hat{\sigma}^+\hat{\sigma}^-$ is bounded above 
whereas $\hat{a}^\dag \hat{a}$ is not. 
Due to these behaviors, $J_{\rm ph}^{\rm out}$ can catch up with $J_{\rm TLS}^{\rm in}$ at certain $\mu_{\rm B}$. 
However, it can not go over $J_{\rm TLS}^{\rm in}$ 
since $J_{\rm ph}^{\rm out} \le J_{\rm TLS}^{\rm in}$ holds in the steady states. 
Therefore the first plateau starts from this value of  $\mu_{\rm B}$. 
As $\mu_{\rm B}$ increases further, $\mu_{\rm B}-\mu$ gets equal to $E_1-E_0$, 
which results in the open of a new channel -- the first excited state -- for pumping from the TLS bath.
Therefore the system can go out of the first plateau when $\mu_{\rm B}-\mu \simeq E_1-E_0$. 
This scenario requires the existence of $J_{\rm ph}^{\rm out}$, 
which is consistent with the fact that the cavity loss is crucial for the plateaus. 

Similarly, the beginning of the second plateau is the point 
where $J_{\rm ph}^{\rm out}$ catches up with $J_{\rm TLS}^{\rm in}$ again. 
In this case, we should note that $\mu_{\rm B}-\mu$ crosses with $E_2-E_1$ before the second plateau starts. 
Thus we do not observe the end of the plateau at this point 
although the pumping channel of the second excited state opens. 
Instead, we observe the effect of this channel open as the jumps 
in $\langle N_{\rm ph} \rangle_{\rm ss}$ [Fig.~\ref{fig:frequency}(a)] and in the currents [Fig.~\ref{fig:frequency}(b)]. 
The end of the second plateau is the point where $\mu_{\rm B}-\mu$ crosses with $E_3-E_2$ 
and the pumping channel of the third excited state opens. 

Finally we remark that  in the plateau regions 
the input excitation of the TLSs is almost perfectly converted to the output of the photons
because $J_{\rm TLS}^{\rm in} \simeq J_{\rm ph}^{\rm out}$ holds there. 
In this sense, conversion efficiency of this system in the plateaus is nearly 100\%. 


{\it Details of the MF-QME}.---The Hamiltonian of the total system (array + baths) is given by 
$\hat{H}_{\rm tot} = \hat{H}_{\rm S} + \hat{H}_{\rm bath}^{\rm TLS} + \hat{H}_{\rm bath}^{\rm ph}$, 
where $\hat{H}_{\rm S}$ is given by Eq.~(\ref{H_S}), and 
$\hat{H}_{\rm bath}^{\rm TLS}$ and $\hat{H}_{\rm bath}^{\rm ph}$ are respectively the Hamiltonians 
of the TLS and photon baths including the system-bath interactions. 
We assume that the baths are sufficiently larger than the array system, 
so that the TLS bath remains in the equilibrium state with $\mu_{\rm B}$ and $\beta$ 
and the photon bath remains in the vacuum. 
We obtain the MF-QME for the array system as follows. 

First, we explain the MF approximation. 
As we mentioned before, we introduce the oscillating photon field as the mean field: 
$\psi(t) \equiv \langle \hat{a}_i \rangle_{\rm ss} = \psi e^{-i\mu t}$. 
Then we use the MF approximation \cite{Greentree_etal}: 
$\hat{a}_i^\dag \hat{a}_j \simeq \psi e^{i\mu t} \hat{a}_j + \psi e^{-i\mu t} \hat{a}_i^\dag - \psi^2$. 
Substituting this into the second term in Eq.~(\ref{H_S}), 
we obtain the MF system Hamiltonian as a sum over the single sites. 
We can thus reduce the problem to the single-site one, and will omit the site index: 
$\hat{H}_{\rm S}^{\rm MF} 
= \hat{H}^{\rm JC} - z\tau(\psi e^{i\mu t} \hat{a} + \psi e^{-i\mu t} \hat{a}^\dag - \psi^2)$, 
where $z$ is the number of the nearest neighbor sites. 
Furthermore, by transforming into the rotating frame with the frequency $\mu$, 
we eliminate the time dependence in $\hat{H}_{\rm S}^{\rm MF}$; 
i.e., $\tilde{H}_{\rm S}^{\rm MF} 
= e^{i\mu\hat{N}_{\rm tot} t} \hat{H}_{\rm S}^{\rm MF} e^{-i\mu\hat{N}_{\rm tot} t}
= \hat{H}^{\rm JC} - z\tau(\psi \hat{a} + \psi \hat{a}^\dag - \psi^2)$, 
where $\hat{N}_{\rm tot} = \hat{N}_{\rm S}+\hat{N}_{\rm bath}$, 
$\hat{N}_{\rm S} = \hat{a}^\dag \hat{a} + \hat{\sigma}^+ \hat{\sigma}^-$ 
is the total number of photon and TLS excitations in the cavity, 
and $\hat{N}_{\rm bath}$ is the excitation number in the baths.
Then we have the von Neumann equation for the density matrix $\tilde{\rho}_{\rm tot}$ 
of the total system in the rotating frame as 
$d\tilde{\rho}_{\rm tot}(t)/dt
= -i \bigl[ \tilde{H}_{\rm S}^{\rm MF} + \hat{H}_{\rm bath}^{\rm TLS} + \hat{H}_{\rm bath}^{\rm ph} 
- \mu \hat{N}_{\rm tot}, \tilde{\rho}_{\rm tot}(t) \bigr]$. 

Next, we explain the QME. 
Starting from the above von Neumann equation, 
we take the trace over the baths and use the Born-Markov approximation
to obtain the QME in the Shr\"odinger picture (within the rotating frame) \cite{BreuerPetruccione}: 
$d\tilde{\rho}/dt = \mathcal{L}\tilde{\rho}$, 
where $\tilde{\rho} = {\rm Tr}_{\rm bath}\tilde{\rho}_{\rm tot}$ 
and $\mathcal{L} = -i\bigl[ \tilde{K}_{\rm S}, ~\bigr] + \mathcal{L}_{\rm ph} + \mathcal{L}_{\rm TLS}$. 
Here $\tilde{K}_{\rm S} = \tilde{H}_{\rm S}^{\rm MF} - \mu \hat{N}_{\rm S}$ 
is the effective system Hamiltonian in the rotating frame. 
The superoperator $\mathcal{L}_{\rm ph}$ associated with the photon bath is given by 
\begin{align}
\mathcal{L}_{\rm ph} \tilde{\rho} = -\frac{\kappa}{2} \bigl( \hat{a}^\dag \hat{a} \tilde{\rho} 
+ \tilde{\rho} \hat{a}^\dag \hat{a} - 2 \hat{a} \tilde{\rho} \hat{a}^\dag\bigr), 
\label{L_diss}
\end{align}
where $\kappa$ is the cavity loss rate. 
To rewrite the superoperator $\mathcal{L}_{\rm TLS}$ associated with the TLS bath into a tractable form, 
we introduce eigenoperators of $\tilde{K}_{\rm S}$ as follows. 
We denote by $\hat{P}(E_q)$ the projection operator onto the eigenspace belonging to $E_q$ (eigenvalue of $\tilde{K}_{\rm S}$).
Then we define eigenoperators: 
$\hat{\sigma}^\pm_\omega 
= \sum_q \hat{P}(E_q \pm \omega) \hat{\sigma}^\pm \hat{P}(E_q)$. 
By using these operators we can write $\mathcal{L}_{\rm TLS}$ as 
$\mathcal{L}_{\rm TLS}=\mathcal{L}_{\rm TLS}^+ + \mathcal{L}_{\rm TLS}^-$ 
\cite{BreuerPetruccione}, 
where 
\begin{align}
& \mathcal{L}_{\rm TLS}^\pm \tilde{\rho} 
\label{L_pump}
\\
&= - \frac{\gamma}{2} \sum_{\omega} f^\pm(\omega) 
\Bigl[ \hat{\sigma}^\mp \hat{\sigma}^\pm_\omega \tilde{\rho} 
+ \tilde{\rho} \hat{\sigma}^\mp_\omega \hat{\sigma}^\pm 
- \hat{\sigma}^\pm \tilde{\rho} \hat{\sigma}^\mp_\omega 
- \hat{\sigma}^\pm_\omega \tilde{\rho} \hat{\sigma}^\mp \Bigr]. 
\notag
\end{align}
Here $\gamma$ is the pumping rate, the $\omega$-sum is over the eigenvalue differences, 
$f^+(\omega)=1/\{1+\exp\beta [\omega - (\mu_B-\mu)]\}$ is the Fermi distribution function, 
and $f^-(\omega) = 1-f^+(\omega)$. 
Note that, if the cavity loss is absent ($\kappa=0$), 
$\mathcal{L}_{\rm TLS}$ leads the array system to the equilibrium state 
with the chemical potential $\mu_{\rm B}$ and inverse temperature $\beta$.

{\it Conclusion}.---We have investigated the nonequilibrium steady states in a coupled cavity QED array system 
that is incoherently pumped by a thermal bath for the TLS and dissipated by the vacuum of the photon. 
We have determined a steady-state phase diagram by employing the photon amplitude $\psi$ as the order parameter. 
In the coherent phase (region with $\psi \neq 0$), we have found that there exist cavity-loss induced plateaus 
where the state of the array system becomes unchanged 
as a function of the chemical potential $\mu_{\rm B}$ of the TLS bath.
We have proposed a mechanism of the plateaus. 
There are two essential elements in the proposed mechanism: 
one is the existence of discrete energy levels (or energy gap) in the system, 
and the other is that the output current to one of the baths can catch up with the input current from the other bath 
(the bounded excitation number of the TLS and unbounded photon number would be the origin in our case). 
Therefore, if our scenario is correct, 
similar plateau phenomena should be observed in other systems which have these elements.

\begin{acknowledgments}
This work was supported by a JSPS Research Fellowship for Young Scientists (No. 24-1112) 
and by the JSPS through the FIRST Program. 
K. K. is also supported by the Project for Developing Innovation Systems of MEXT. 
\end{acknowledgments}

\clearpage

\setcounter{section}{1}
\setcounter{equation}{0}
\setcounter{figure}{0}
\renewcommand{\thesection}{\Alph{section}}
\renewcommand{\thefigure}{\Alph{figure}}
\numberwithin{equation}{section}

\section{Supplemental Material for ``Cavity-loss induced plateau in coupled cavity QED array''}
\subsection{Cavity-loss dependence of the beginning of the first plateau}

Here we investigate the dependence of the first plateau on the cavity loss $\kappa$ 
to obtain data that support the essentiality of the cavity loss for the plateaus. 
We denote the chemical potential at the beginning of the first plateau by $\mu_{\rm B}^{\rm 1st}$, 
which is pointed with the up arrow in Fig.~3(a). 

\begin{figure}[ht]
\begin{center}
\includegraphics[width=\linewidth]{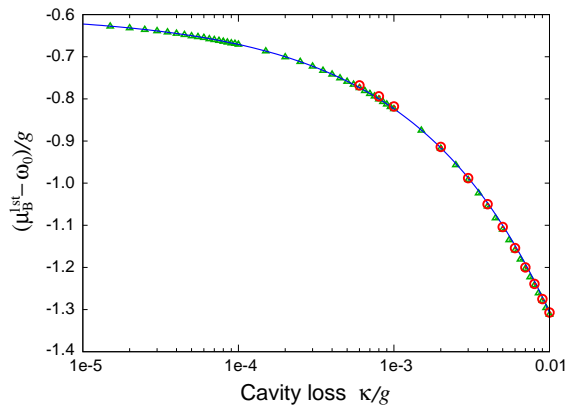}
\end{center}
\caption{Chemical potential $\mu_{\rm B}^{\rm 1st}$ at the beginning of the first plateau (circle symbols) 
as a function of the cavity loss $\kappa$. 
The cutoff for the maximum photon number in the cavity is set to 15 for $\kappa/g \ge 0.002$, 
25 for $\kappa/g=0.001, 0.0008$, and 30 for $\kappa/g=0.0006$. 
Evaluation from the equilibrium state (with the cutoff photon number being 500) is plotted with the triangle symbols. 
The solid line is a plot of Eq.~(\ref{muB1st}). 
The hopping constant $z\tau/g$ is set to 0.6. 
The other parameter values are the same as those in Fig.~2. 
}
\label{fig:muB1st}
\end{figure}

We plot $\mu_{\rm B}^{\rm 1st}$ (circle symbols) as a function of $\kappa$ in Fig.~\ref{fig:muB1st}. 
We observe that $\mu_{\rm B}^{\rm 1st}$ increases as $\kappa$ becomes smaller. 
This is consistent with the essentiality of the cavity loss for the plateaus.

We also evaluate $\mu_{\rm B}^{\rm 1st}$ from the equilibrium state as follows. 
As we mentioned in the main text, we can identify $\mu_{\rm B}^{\rm 1st}$ with the point where 
$J_{\rm ph}^{\rm out}$ becomes equal to $J_{\rm TLS}^{\rm in}$. 
Since the system is near to an equilibrium state for $\mu_{\rm B} < \mu_{\rm B}^{\rm 1st}$, 
we expect that we can calculate $J_{\rm ph}^{\rm out}$ and $J_{\rm TLS}^{\rm in}$ 
for $\mu_{\rm B} < \mu_{\rm B}^{\rm 1st}$ in good approximation by using the equilibrium state. 
We thus evaluate $\mu_{\rm B}^{\rm 1st}$ from the approximate $J_{\rm ph}^{\rm out}$ and $J_{\rm TLS}^{\rm in}$. 
For the practical calculation, we use the ground state $|E_0\rangle$ as the equilibrium state 
because the temperature of the TLS bath ($1/\beta g=0.001$) is sufficiently lower than the eigenenergy difference 
[$(E_1-E_0)/g>0.3$; see Fig.~3(c)]. 
In this case we have $J_{\rm ph}^{\rm out} = \kappa \langle E_0 | \hat{a}^\dag \hat{a} | E_0 \rangle$ 
and $J_{\rm TLS}^{\rm in} = \gamma |\langle E_0 | \hat{\sigma}^- | E_0 \rangle|^2$. 
We note that we set $\mu_{\rm B}-\mu$ to $+0$ to obtain the latter. 
In Fig.~\ref{fig:muB1st}, we plot $\mu_{\rm B}$ (triangle symbols) 
where the thus-evaluated $J_{\rm ph}^{\rm out}$ and $J_{\rm TLS}^{\rm in}$ get equal. 
We see that the result provides a good approximation to $\mu_{\rm B}^{\rm 1st}$. 
We also observe that the approximate $\mu_{\rm B}^{\rm 1st}$ approaches to a finite value as $\kappa$ decreases. 

To see this more clearly, we use another evaluation of $\mu_{\rm B}^{\rm 1st}$ in the following analysis.
First we note that when we approximate the steady state by the ground state 
we have $J_{\rm ph}^{\rm out} = \kappa \langle \hat{a}^\dag \hat{a} \rangle$ 
and $J_{\rm TLS}^{\rm in} = \gamma |\langle \hat{\sigma}^- \rangle|^2$. 
Next, from the steady-state equation for $\langle \hat{a} \rangle$, 
we have a relationship $\langle \hat{\sigma}^- \rangle 
= -(\omega_0 - \mu -z\tau - i\kappa/2) \langle \hat{a} \rangle/g$. 
Substituting this relationship into $J_{\rm TLS}^{\rm in}$ 
and equating $J_{\rm TLS}^{\rm in}$ with $J_{\rm ph}^{\rm out}$ (with $\mu=\mu_{\rm B}^{\rm 1st}$), we have 
$|\omega_0 - \mu_{\rm B}^{\rm 1st} -z\tau - i\kappa/2|^2 
= \kappa \langle \hat{a}^\dag \hat{a} \rangle/\gamma |\langle \hat{a} \rangle|^2$. 
Now we neglect $O(\kappa^2)$ term (since we consider small $\kappa$) 
and approximate $\langle \hat{a}^\dag \hat{a} \rangle \simeq |\langle \hat{a} \rangle|^2$
(since we consider the coherent phase). 
Then we obtain 
\begin{align}
(\mu_{\rm B}^{\rm 1st} - \omega_0)/g = -z\tau/g - \sqrt{\kappa/\gamma}.
\label{muB1st}
\end{align}
We plot this equation as the solid curve in Fig.~\ref{fig:muB1st}. 
We observe that this equation provides a good approximation to $\mu_{\rm B}^{\rm 1st}$. 
Equation (\ref{muB1st}) means that $(\mu_{\rm B}^{\rm 1st} - \omega_0)/g$ converges to $-z\tau/g$ 
in the limit of $\kappa \to 0$. 
At first glance this may look inconsistent with the essentiality of the cavity loss for the plateaus. 
However, in the equilibrium case, the region with $(\mu_{\rm B} - \omega_0)/g \ge -z\tau/g$ is unstable 
\cite{KochHurSupplement}. 
Therefore we conclude that the plateaus do not appear in the equilibrium situation ($\kappa=0$).
We thus show that the results in this supplement support the essentiality.


\begin{thebibliography}{99}
\bibitem{note1}
In the incoherent phase, the photon frequency $\mu$ is indefinite because the photon amplitude $\psi$ vanishes.
\bibitem{note2}
There is no current from the photon bath to system because the photon bath is in the vacuum state.
\end{thebibliography}
\end{document}